\documentclass[letterpaper, 10 pt, conference]{ieeeconf}  

\usepackage{graphicx}

\IEEEoverridecommandlockouts                              
\title{\LARGE \bf
Technical Report on Intruder Detection and Alert System
}

\author{Manish Kumar, Shubham Kaul
\thanks{$^{1}$Manish Kumar, $^{1}$Shubham Kaul are with the Department of Electronics and Communication Engineering, IIIT-Delhi, INDIA
{\tt\small  manish12142@iiitd.ac.in, shubham12161@iiitd.ac.in \newline }}%
\thanks{$^{2}$This project was guided by Dr. Vivek Ashok Bohara who is the faculty of Department of Electronics and Communication Engineering, IIIT-Delhi, INDIA
        {\tt\small vivek.b@iiitd.ac.in}}%
}

\begin{document}

\maketitle
\thispagestyle{empty}
\pagestyle{empty}

\begin{abstract}

This work presents a smart trespasser detection and alert system which aims to increase the amount of security as well as the likelihood of positively identifying or stopping trespassers and intruders as compared to other commonly deployed home security system. Using multiple sensors, this system can gauge the extent of danger exhibited by a person or animal in or around the home premises, and can forward certain critical information regarding the same to home owners as well as other specified persons such as relevant security authorities.

\end{abstract}


\begin{keywords}
Home security, wireless device network, Zigbee, WiFi
\end{keywords}
\section{INTRODUCTION}

Various security systems can be found installed in homes and offices alike. For example, Closed-circuit Television (or CCTV) is a popular technology used extensively for the purpose of home security and surveillance. However, while such systems may offer some merits in the form of low cost and relative ease of installation, they also present some major drawbacks. Such systems usually have no mechanism to send critical information to users/home owners if they are not present in the home premises. Additionally, since it records video continuously and stores it locally, one has to scroll through long durations of time, often hours in order to get to the relevant section of the video. 
The system presented in this paper aims to improve upon such shortcomings of other security systems. Hence, this system can send relevant video data to the users and home owners directly, without the need of an additional centralized monitoring office to relay data from. Since all of this is done in real time, this data could also be specified to be sent to security authorities which would ideally expedite the process of identification or capturing of an intruder. The system [6] is implemented in such a way that it could be integrated with already installed Home Automation System like, [2] and [4] very easily.

\section{Implementation/Key Features}

By using a combination of sensors and microprocessors that are described in the following sections, this system is able to incorporate some crucial features into its functioning. These are explained below:
\begin{itemize}
\item This system uses sensors to monitor the presence of trespassers and possible intruders near the home’s entry points such as door, windows etc. This enables the system to start recording video only when it “perceives danger”. Hence, a user does not have to scan through hours of irrelevant video data. This not only saves time but also decreases the overall memory and processing power requirements.

\item This system uses the home’s wireless LAN network to send an e-mail to the user with this video as an attachment. No additional network hardware installations are required in doing so.

\item This system also uses a sensor placed on the inside of the home premises, near doors/windows etc in order to detect the intrusions and sends another mail immediately to the home owner as well as security authorities in this case.

\begin{figure}[h!]
\centering
\includegraphics[width=0.45\textwidth]{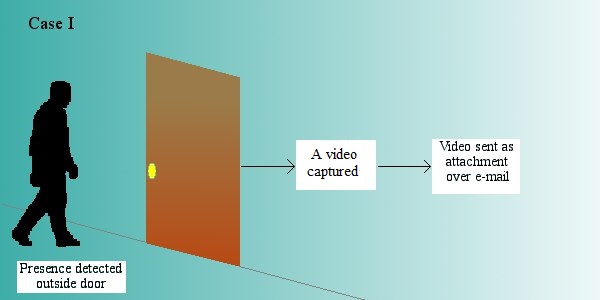}
\caption{\label{fig:Presence}System response when presence detected near door}
\end{figure}

\begin{figure}[h!]
\centering
	\includegraphics[width=0.45\textwidth]{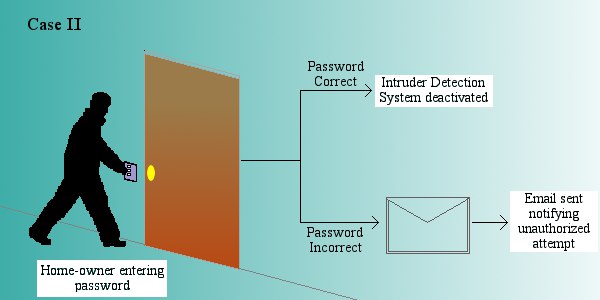}
\caption{\label{fig:PM}System response when password mechanism is used}
\end{figure}

\begin{figure}[h!]
\centering
\includegraphics[width=0.45\textwidth]{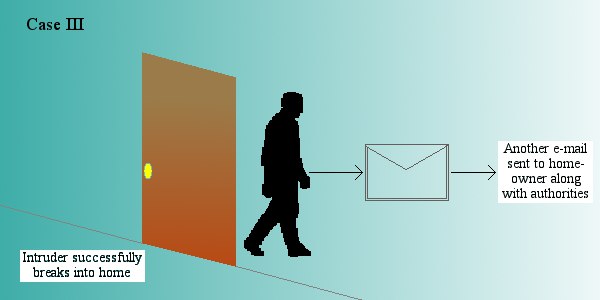}
\caption{\label{fig:Intrusion}System response in case of a break-in}
\end{figure}

\item  Since this system works on wireless protocols such as IEEE 802.11 [3] and IEEE 802.15.4 [1] (explained in later sections), it could also be extended to interact with other devices present in the home premises during an intrusion in order to maximize security.

\item  Since the home owners or other authorized members would not want to trigger the security alarm system, there is also an integrated deactivating mechanism placed outside the door, implemented in the system. It presents an alternative password input mechanism to traditional PIN-based password systems (explained in section IV).

\end{itemize}

\section{Hardware Setup}

For implementing the previously mentioned features, three processing units have been used- a microcontroller, a raspberry pi single board computer and a PC (with LabVIEW installed)

\begin{itemize}

\item The microcontroller [9] is the unit which wirelessly sends an intruder alert to the PC in case of a break-in. To do this, the microcontroller [9] uses a combination of a Laser and an LDR as input, to detect an unauthorized opening of the door.

\begin{figure}[h!]
\centering
\includegraphics[width=0.48\textwidth]{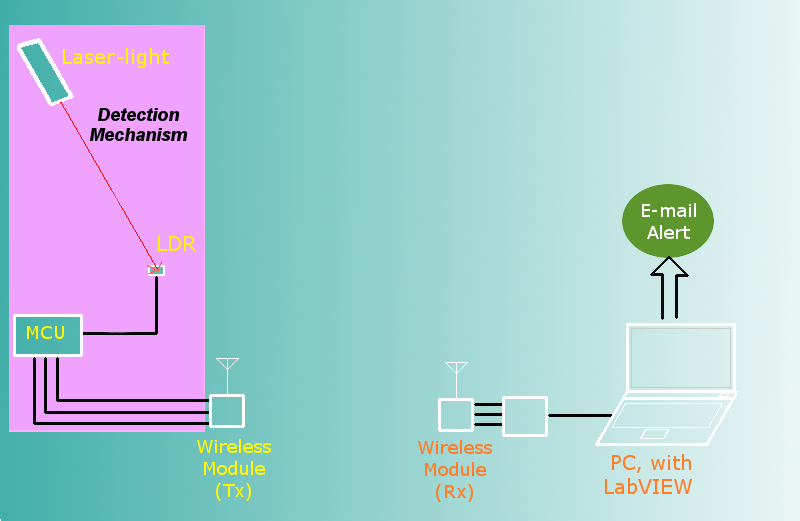}
\caption{\label{fig:ipfinal}Hardware Setup inside the house}
\end{figure}

\begin{figure}[h!]
\centering
\includegraphics[width=0.48\textwidth]{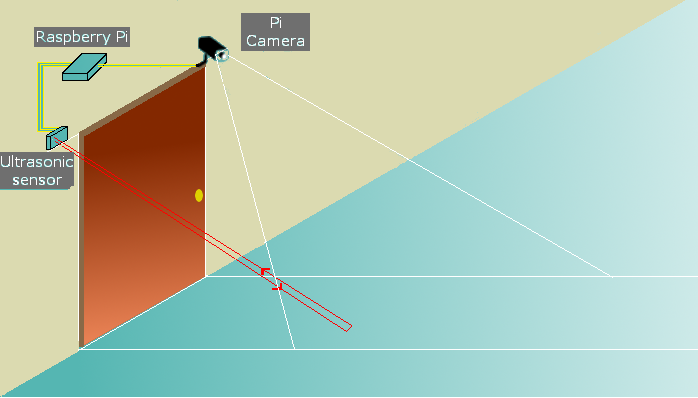}
\caption{\label{fig:outdoor2}Hardware Setup outside the house}
\end{figure}

Any kind of obstruction of this laser beam triggers the wireless module attached with microcontroller to send the intruder alert. The deactivation mechanism for this system is also written within the microcontroller. The input console for the deactivation mechanism is placed intelligently outside the door.   
\item The PC, after receiving an intruder alert from the microcontroller[9] sends an email using the software installed in the system (which uses LabVIEW [10] API) to the home-owner as well as authorities regarding the intrusion. 
\item The Raspberry Pi Single Board Computer [5] on the other hand handles the image/video capturing aspect of the project. The Raspberry Pi single board computer uses an Ultrasonic sensor to detect the presence of an individual near the door. After this sensor is triggered by the presence of a person, it starts recording a brief 5-10 second video of that area, in order to capture the identity of the trespasser or any activity happening outside the door on footage. This video is then both saved on-board and sent to the home owner and/or authorities via e-mail. 
\end{itemize}

\section{Deactivation Mechanism}
In this system, a new type of password lock mechanism has been implemented as an alternative to the traditional numeric password input mechanism. Users have to be trained to input the password into the deactivation mechanism, thereby further increasing the security of the system. The users are given a sequence of 1s and 0s (or a “press” and “don’t press”) that essentially serve as their password. In order to correctly input this password into the system, the user presses a button, which initiates the password-entering mode. Once this mode is entered, an LED blinks periodically for n times. While the LED is temporarily on in each of its n iterations, the user has to press or not press the password button. This is determined by what password sequence has been given to the user. Hence, if a user has, say, a 7-bit-long password given by “1100101”, he/she would press the password button only on the 1st, 2nd, 5th and 7th LED pulse, and not on the rest.
This mechanism is illustrated in the Fig 6. and Fig 7. The upper waveforms represent the expected pulse sequence by the system i.e. the waveform corresponding to the preset password of the system. In Fig. 6 the user enters a sequence that does not match with the expected sequence, and hence, the system is not deactivated. Additionally, since an incorrect password entry may resemble a potential threat (by an authorized person trying to gain access), the system e-mails the home-owner regarding this unsuccessful deactivation attempt. In Fig. 7, the pulses generated by the home owner on pressing the password button correspond exactly to that of the password, and hence, the system is deactivated. An e-mail is sent here too, for additional security purposes, alerting users of a successful deactivation. The system would now be disabled and would not send a mail in case the door is opened.

\begin{figure}[h!]
\centering
\includegraphics[width=0.48\textwidth]{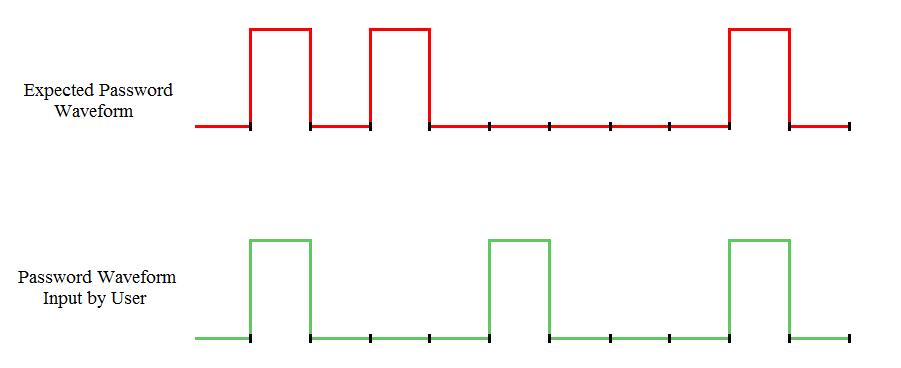}
\caption{\label{fig:ipfinal}When a user inputs an incorrect password}
\end{figure}

\begin{figure}[h!]
\centering
\includegraphics[width=0.48\textwidth]{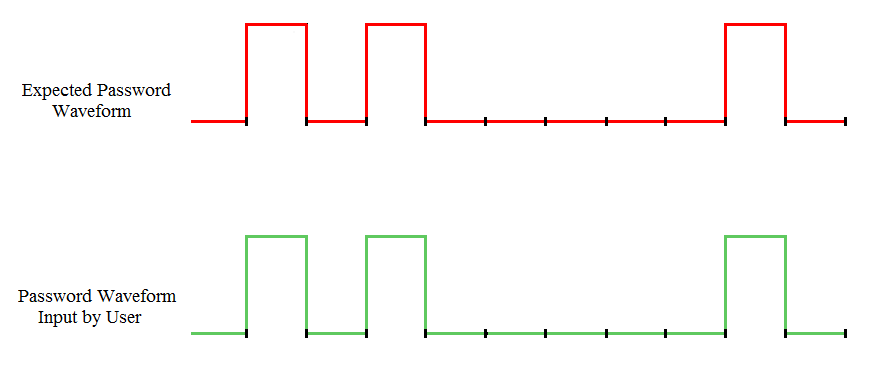}
\caption{\label{fig:ipfinal}When a user inputs the correct password}
\end{figure}

\section{Components and Technologies Used}
\begin{table}[h!] [h!]
\centering
\begin{tabular}{|l|p{2 cm}|p{2 cm}|l|} \hline
S.No. &  Component & Product Code & Quantity \\ 
\hline
1 & Xbee Radio Series 2 Zigbee Module by Digi International Inc. [1] & XB24-Z7WIT-004 revK & 2\\\hline
2 [1] & Arduino Microcontroller Development [8] Board & Deumilanove using ATMega328p [9] & 1 \\\hline
3 & Laser Light & Commonly Available & 1 \\\hline
4 & Light-Dependent Resistor & Commonly Available & 1 \\\hline
5 & PC, with LabVIEW & LabVIEW version 14.0 by National Instruments & 1 \\\hline
6 & Zigbee Explorer USB Board & XUAB-ZIGBEE BTBEE & 1\\\hline
7 & Zigbee Adapter & NR-RF-07 & 1 \\\hline
8 & Raspberry Pi Single Board Computer [5] & Raspberry Pi 1 - Model B & 1 \\\hline
9 & Raspberry Pi Camera Module & [5] 5MP Pi Camera & 1 \\\hline
10 & Ultrasonic Sensor & HC-SR04 [7] & 1\\\hline
\end{tabular}
\caption{\label{tab:component}Components Used}
\end{table}

\subsection{Zigbee Wireless Transceiver Module (Series 2)}
Zigbee is a low cost mesh network specification that is used for creating Wireless Personal Area Network built from the small, low power digital radios. It is based on the IEEE 802.15.4 [1] standard and used to create wireless networks which require low data rates, energy efficiency and  secured communication. Due to low power consumption, it covers the area(range) from 10 to 100 metres. The range of the network can be extended by creating a mesh network of several digital radios. If we compare technology defined by Zigbee specification, it uses lesser power than other wireless personal area networks such as Bluetooth, Wi-fi etc. Zigbee operates in different radio bands, 2.4GHz in most countries, 868 MHz in Europe, 784 MHz in China and 915 MHz in USA and Australia. Hence, data rates vary from 20 kbit/s to 250 kbit/s [1]. Zigbee chips are integrated with the radios which further used in the formation of low cost wireless personal networks that require low power consumption and communicate with low data rates [1]. So, there are two types of XBee radio physical hardware that are based on Zigbee:
\subsubsection{XBee Series 1 hardware}
A microchip made by Freescale is integrated with these radios to provide low cost, low power consumption, simple, low data rates and zigbee standard based point-to-point communications, which further can also be used in implementation of mesh networking [1].
\subsubsection{XBee Series 2 hardware}
The functionality of XBee Series 2 radio is same as Series 1, it is just an updated version of Series 1 which uses a microchip from Ember Networks instead of Freescale, because of which it provides several different flavors of standards-based ZigBee mesh networking. Mesh networking comes in very handy while creating optimum sensor networks [1].
\begin{figure}[h!]
\centering
\includegraphics[width=0.48\textwidth]{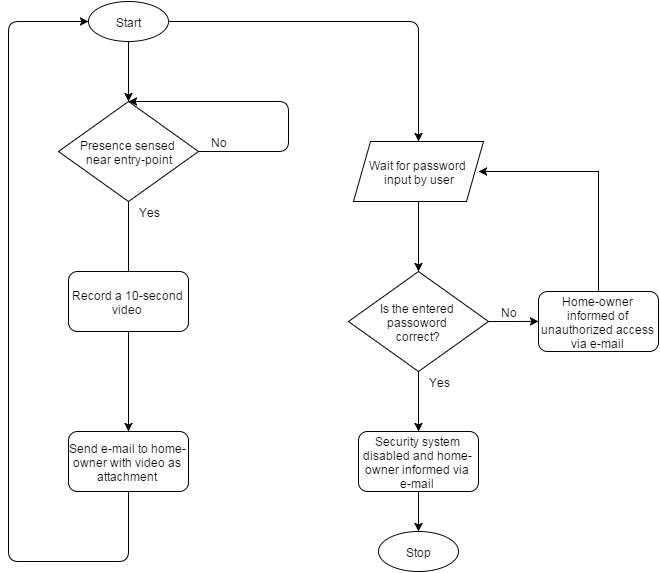}
\caption{\label{fig:ipfinal}Flow graph of mechanism outside door}
\end{figure}

Series 2 hardware provides the access to full ZigBee protocol [1]. Series 1 is optimal for smaller-sized wireless systems whereas Series 2 is designed so as to support larger sensor networks also and is ideal for the precise communication with other ZigBee standards-based systems that are now very popular and being widely implemented for residential, academic, and industrial purposes [1]. The Series 2 hardware offers greater features as compared to Series 1 hardware such as better range and better power efficiency. Both hardware follow the same wireless protocol to form the WPANs and can be easily interchanged in any network, with only few changes to be made in the software. However, the point to be noted here is that Series 2 hardware will not communicate with the Series 1 hardware at all [1]. Each wireless network must use either Series 1 or Series 2 hardware. 

\subsection{Arduino Microcontroller Development Board}
Arduino [8] is a programmable electronic development board that uses microcontroller chips as the CPUs. This microcontroller forms the hardware circuit board, whereas the Arduino IDE [8] shipped along with such boards are the software used on the PC to write and upload codes from. The microcontroller chip used in the Arduino Board used in this project is the ATMega 328p chip [9] - a single chip micro-controller created by Atmel and belonging to the megaAVR series. The Arduino Development Boards [8] use standard connectors which allows connecting the CPU board to a variety of interchangeable add-on modules known as shields. For the purposes of this project, we have connected the wireless transceiver shield explained in Section IV.A.

\begin{figure}[h!]
\centering
\includegraphics[width=0.48\textwidth]{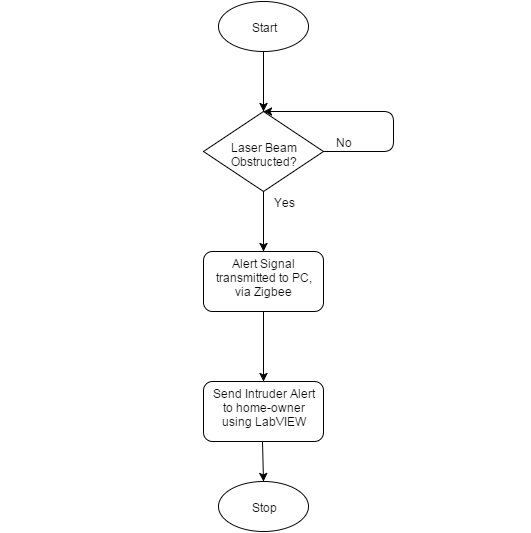}
\caption{\label{fig:ipfinal}Flowgraph of mechanism inside the door}
\end{figure}

\subsection{Light Dependent Resistor}
A light-dependent resistor (LDR), also known as a photoresistor, is an electronic passive circuit element, that behaves as a light-controlled variable resistance. An LDR offers decreased resistance with increasing incident light intensity; thus exhibiting photoconductivity.

\subsection{Zigbee Explorer USB Board}
A Zigbee Explorer USB is a hardware component that allows a Zigbee module (such as the one elaborated upon in Section IV.A) to be interface with a PC. The explorer board contains the female connectors of the zigbee wireless module pins. On the other end, it has a male connector for the USB COM port to be connected into the USB socket on a PC. Any program monitoring that particular COM port can then receive or analyse the signals received by the connected Zigbee Module, or even send signals via the same channel.
\subsection{Zigbee Adapter}
A Zigbee Adapter is a hardware component that helps in connecting a Zigbee Module (similar to the one described in Section IV.A ) to an embedded device (in our project, to an Arduino similar to the one described in Section IV.B) with digital/electronic pins. This enables such devices to transmit signals wirelessly via the wireless module, using the read/write pins to interface with the said module.

\subsection{Raspberry Pi Single Board Computer}
Raspberry Pi computers [5] are small computers developed by the Raspberry Pi Foundation. These feature ports and connectors similar to the commonly used home and office desktop computers (such as Ethernet Port, USB ports, HDMI port and even a 3.5mm audio socket) but the whole operating unit is contained within a much smaller, credit-card-sized board [5]. When connected to peripherals such as a monitor  or TV screen, keyboard and mouse via the above mentioned connectors, it can be used in a fashion similar to regular desktop PCs (albeit being less powerful and running lighter operating system). Additionally, this single-board computer also features certain digital pins that can be used for general purpose interfacing with electronic components [5] (in our case, the Ultrasonic sensor, which has been explained in Section IV.H)

\begin{figure}[h!]
\centering
\includegraphics[width=0.48\textwidth]{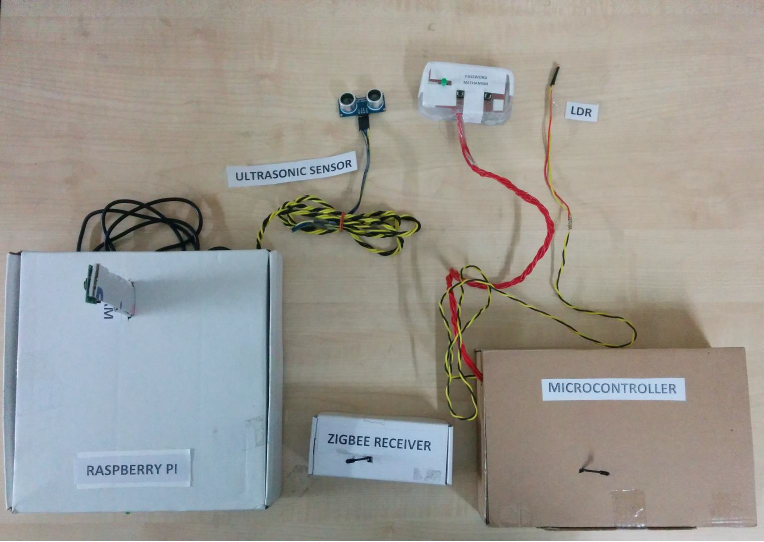}
\caption{\label{fig:ipfinal}Hardware after packaging}
\end{figure}

\subsection{Raspberry Pi Camera Module} 
The Raspberry Pi Foundation also manufactures a standard add-on camera module to be used with their Raspberry Pi [5] computers. This camera has a five megapixel resolution and supports clicking both still images and videos (1080p30, 720p60 and VGA90 modes) this camera module is connected to the Raspberry Pi on its CSI port via a 15cm ribbon cable [5].

\subsection{Ultrasonic Sensor}
Ultrasonic Sensors are specifically made to sense the object proximity or range with the help of ultrasound reflection, just like a radar, which helps in calculating the time taken to reflect ultrasound waves between the sensor and a solid object [7]. It consists of three units, one or more transmitters, receiver and control unit. The transmitters emit a high frequency ultrasonic sound waves which reflect back from any nearby solid object and then received by the receiver and further processed by the control unit to calculate the time taken. Using this time, the distance between the object and the sensor can be calculated using some calculations. Thus this distance can be used for specific purposes. In this project, we are using this sensor to capture any motion within a specific range. We have used "HC-SR04" ultrasonic sensor [7] in this project. It has four pins: 5V power supply (Vcc), Trigger Pulse Input (TRIG), Echo Pulse Output (ECHO) and ground (GND) [7]. Through raspberry pi, we send an input signal to TRIG which then triggers the sensor to send ultrasonic pulse. ECHO remains low until sensor is triggered. When the sensor receives the waves it measures the time and then sends a 5V signal to ECHO. We have used a voltage divider to feed the signal coming from ECHO to the GPIO pin of raspberry pi as it is rated at 3.3V but ECHO is rated at 5V [7]. After this sensor detects a distance reduced more than a particular threshold distance, as specified in the code (i.e. due to an obstruction in the path), it sends a high bit to the attached Raspberry Pi [5].

\section{Future Work}
Further, this project can be easily expanded to include various additional features, functionalities and services. For example, the password/deactivation mechanism could be replaced from the current button-input style to a traditional PIN based system or even a voice-command based activation/deactivation of certain part of this security system.
Also, in order to completely monitor all activity near the home premises, multiple Raspberry Pi [5] computers can be integrated instead of just one. Such Raspberry Pi computers would be located outside possible entry points into the house such as windows, vents, etc., and not just the main door. Since the Raspberry Pi single board computers operate independently of the microcontroller based unit in our implementation, such expansion would be extremely easy.
Lastly, the IEEE 802.15.4 [1] protocol is immensely used for inter-device communication and Internet-of-Things implementations. Hence, the number of ways in which this system could be used in conjunction with other IEEE 802.15.4 enabled devices are endless [1]. In cases where the home owner may not be connected to the internet, he/she would not be able to check the emails sent by this security system. To solve this issue, an SMS sending feature can also be integrated in the current system which will immediately notify the owner and concerned authorities about the intrusion/unauthorized access.

\section{Acknowledgement}
We are thankful to Vivek Ashok Bohara, ( Assistant Professor, Department of Electronics and Communication Engineering, IIIT-Delhi, India ) for all possible guidance. We are also thankful to Mr. Vibhutesh Kumar Singh ( M.Tech. Student, Department of Electronics and Communication Engineering, IIIT-Delhi, India ) for, LabVIEW [10] logic development, his valuable assistance while planning and executing throughout the project. This work is a part of Project iDART- Intruder Alert and Detection in Real Time, which is submitted as en entry to CII Innovation Challenge, 2015, India, as well as in NI Engineering Impact 2015.

\addtolength{\textheight}{-12cm}   



\end{document}